\documentclass[journal]{IEEEtran}
\IEEEoverridecommandlockouts
\usepackage{cite}
\usepackage{amsmath,amssymb,amsfonts}
\usepackage{algorithmic}
\usepackage{graphicx}
\usepackage{textcomp}
\usepackage{xcolor}
\usepackage{soul}
\usepackage{pgfplots}
\usepackage{pgfplotstable}
\pgfplotsset{compat=1.18}
\usepgfplotslibrary{statistics}
\pgfplotsset{grid style={dotted,gray}}
\usepackage{tikz}
\usepackage{caption}
\usepgfplotslibrary{groupplots}
\usepackage{comment}
\usepackage{booktabs}
\usepackage{colortbl}
\definecolor{UEPdark}{RGB}{180,30,30}
\definecolor{UEPlight}{RGB}{230,120,120}
\definecolor{EEPdark}{RGB}{30,80,180}
\definecolor{EEPlight}{RGB}{120,170,230}
\usepackage{soul}
\usepackage{tikz}
\usepgfplotslibrary{fillbetween}
\usepackage{comment}
\usepgfplotslibrary{groupplots}
\definecolor{UEPdark}{RGB}{180,30,30}   % lambda = 0.7
\definecolor{UEPlight}{RGB}{230,120,120} % lambda = 0.9
\definecolor{EEPdark}{RGB}{30,80,180}    % dark blue
\definecolor{EEPlight}{RGB}{120,170,230} % light blue

\def\BibTeX{{\rm B\kern-.05em{\sc i\kern-.025em b}\kern-.08em
    T\kern-.1667em\lower.7ex\hbox{E}\kern-.125emX}}
\begin{document}

\title{Adaptive Unequal Error Protection for Semantic Split Learning over Wireless Channels}

\author{Vukan Ninkovic, ~\IEEEmembership{Member,~IEEE,} Dejan Vukobratovic,
~\IEEEmembership{Senior Member,~IEEE,} Dragisa Miskovic,
~\IEEEmembership{Member,~IEEE,} Chao Wang, ~\IEEEmembership{Member,~IEEE}
\thanks{V. Ninkovic is with the University of Novi Sad, Serbia, and the
Institute for Artificial Intelligence Research and Development of Serbia
(e-mail: ninkovic@uns.ac.rs); D. Vukobratovic is with the University of
Novi Sad, Serbia; D. Miskovic is with the Institute for Artificial
Intelligence Research and Development of Serbia; C. Wang is with the College
of Electronic and Information Engineering, Tongji University, Shanghai,
China.}
\thanks{This work is supported by the Serbian Ministry of Science,
Technological Development and Innovation (Serbia–China Cooperation Project
No. 00101957 2025 13440 003 000 620 021) and Intergovernmental International
Science and Technology Innovation Cooperation of National Key Research \&
Development Program of China under Grant 2024YFE0197400.}
}

\maketitle

\begin{abstract}
We propose a task-aware semantic split learning (SL) framework for wireless edge--cloud inference, in which the reliability of transmitted latent representations is dynamically adapted to their relevance for the downstream task.
An autoencoder (AE)-based physical (PHY) layer enables end-to-end learning of the communication interface, while unequal error protection (UEP) is realized via mutual information (MI)-driven prioritization of latent components during training.
The gradient of the estimated MI with respect to each latent component serves as a sensitivity-based proxy for task relevance, providing a fully learning-driven prioritization that adapts to both the data distribution and the downstream task.
We further show that this prioritization translates into measurable physical-layer effects: MI-guided UEP assigns significantly higher transmit power to the most task-critical latent components compared to the equal error protection (EEP) baseline.
Experiments on real-world IoT sensing data demonstrate consistent gains over equal and fixed-UEP baselines across SNR regimes.
Additional analysis confirms ranking stability, estimator robustness and 
 generalization across datasets and task types, indicating broad applicability 
of the proposed framework.
\end{abstract}

%\Rone{Additional experiments confirm the stability of the learned ranking across batch sizes and its generalization to a second dataset.}


\section{Introduction}

Wireless IoT inference systems increasingly rely on transmitting intermediate representations rather than raw data, requiring communication mechanisms that prioritize information relevant for the downstream task under strict bandwidth and reliability constraints~\cite{saad_2020}.
Semantic communication improves efficiency and robustness under such constraints by focusing on task-relevant representations~\cite{beyond_2023,weaver_1953,papas_2021}.

Split learning (SL) distributes intelligence across edge devices and the cloud~\cite{gupta_2018}. In SL, a lightweight model at the edge extracts intermediate features from raw data and sends them to the server for further inference. This reduces edge computation while establishing a semantic interface whose reliability directly impacts task performance. 

At the physical (PHY) layer, autoencoder (AE)--based designs replace conventional modulation and coding with end-to-end trainable neural transceivers that adapt to noise and fading~\cite{OShea_2017}. Such designs naturally enable unequal error protection (UEP), allowing different reliability levels to be assigned to different parts of the transmitted representation~\cite{he_2024,ninkovic_2025}.

Recent work has explored semantic communication with AE-based PHY design and distributed learning~\cite{gunduz_2023,he_2024}, including information bottleneck approaches~\cite{RMIB_ref,dv_bound}. Recent semantic UEP schemes prioritize tokens or features by semantic importance~\cite{zhang_tokcom,zhou_dualimportance, Liu_2026}, but do not derive UEP from the task relevance of individual latent components at the split-learning interface. In contrast, our approach uses the downstream task to identify the latent components that require stronger protection, enabling task-aware reliability allocation directly at the semantic split.

The key novelty lies in reformulating UEP at the SL interface as a \emph{task-driven information allocation problem}. Rather than relying on predefined protection classes or reconstruction-based criteria, protection levels are determined dynamically based on the contribution of each latent component to the prediction task, quantified via its mutual information (MI) with the target variable.
The contributions are:
(1) We develop a semantic SL framework in which intermediate representations are transmitted over a fading wireless channel using an AE-based PHY layer;
(2) We propose a task-aware ranking criterion that dynamically prioritizes latent components based on MI sensitivity during end-to-end training;
(3) We demonstrate that the learned prioritization induces PHY-layer reliability differentiation, allocating higher transmit power to task-critical components and improving inference performance over equal error protection (EEP) and fixed-UEP baselines.

\section{Background \& System Framework}
\subsection{Theoretical Background}

\begin{figure}[!t]
    \centering
    \captionsetup{justification=centering}
    \includegraphics[width=0.7\linewidth]{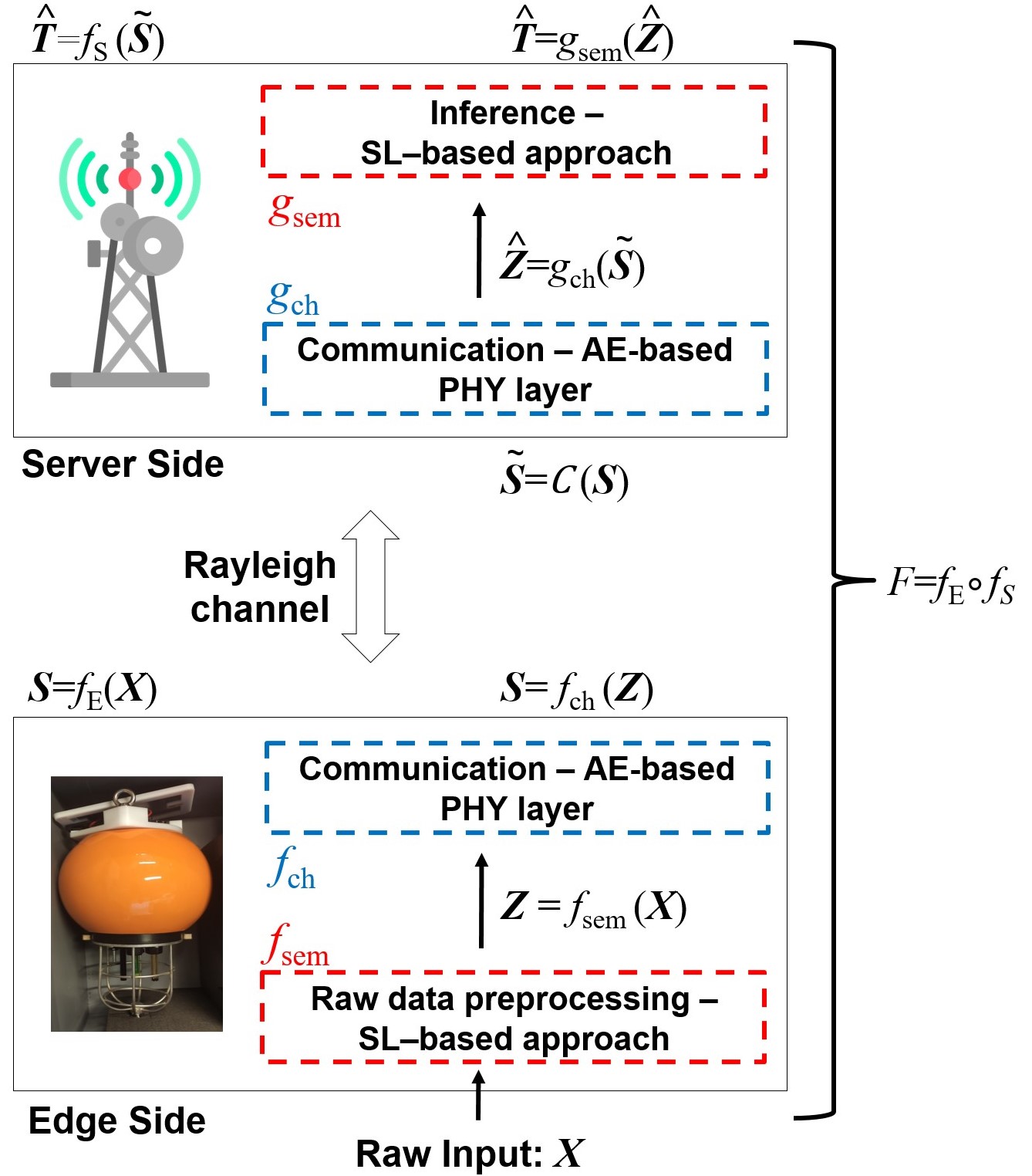}
    \caption{SL-aided semantic communication architecture.}
    \label{fig:sys_mod}
\end{figure}

\textbf{SL for edge--cloud inference:}
SL partitions a neural network $F$  between an edge device and a
central server~\cite{gupta_2018}.
 Let $X \in \mathbb{R}^{N}$  be the edge input and $T(X)$ the
 task output.
 The overall model $F$ is decomposed as
$F = f_{\mathrm{S}} \circ f_{\mathrm{E}}$ (Fig.~\ref{fig:sys_mod}), where
$f_{\mathrm{E}} : \mathbb{R}^{N} \rightarrow \mathbb{R}^{n}$
 and
$f_{\mathrm{S}} : \mathbb{R}^{n} \rightarrow \mathbb{R}$. %\footnote{While the framework supports arbitrary output dimensions, in this
%work the task reduces to predicting a single real-valued scalar.}.
 The edge device computes $S = (s_1,..,s_n)=  f_{\mathrm{E}}(X)$, $S \in \mathbb{R}^{n}$,
and the server produces an estimate $\hat{T} = f_{\mathrm{S}}(\tilde{S})$, as illustrated in Fig.~\ref{fig:sys_mod}.
By separating feature extraction from final inference, SL reduces
communication overhead: the centralized baseline requires transmitting the full
input of dimension $N$, whereas the proposed framework transmits a compressed
latent representation of dimension $n \ll N$ (see Section~\ref{sec:performance_comparison}).

\textbf{Semantic communication paradigm:}
In contrast to conventional communication systems, the receiver aims to generate a task-relevant output $\hat{T} \approx T(X)$ directly from the received representation, without requiring faithful reconstruction of the original input~\cite{weaver_1953,papas_2021}. This objective can be formalized through the Information Bottleneck (IB) principle~\cite{tishby_2000}, which seeks a representation $Z=(z_1,\ldots,z_K)$, $Z\in \mathbb{R}^K$ that preserves task-relevant information about $T$ while minimizing redundancy with the input:
\begin{equation}
\min I(X;Z) \quad \text{s.t.} \quad I(Z;T(X)) \geq \epsilon,
\end{equation}
where $I(\cdot;\cdot)$ denotes MI and $\epsilon$ sets the required task
fidelity.
The key observation that motivates this work is that minimizing
reconstruction error does not guarantee preservation of the sufficient
statistics of $X$ for predicting $T$~\cite{fayaz_2024}: a representation may
achieve high reconstruction fidelity while discarding features critical for the
downstream task, leading to the divergence between reconstruction MSE and
task-level accuracy reported in Section~\ref{sec:performance_comparison}. Consequently, unlike IB/variational-IB methods~\cite{tishby_2000}, which determine \emph{what} information to retain in $Z$, the proposed framework addresses \emph{how} transmission reliability should be allocated across the learned latent representation.

\textbf{AE-Based PHY design with learning-driven UEP:}
AE-based approaches  replace conventional modulation and coding  with
an end-to-end trainable system~\cite{OShea_2017}.
Here, UEP can be captured by a weighted reconstruction objective applied in the latent space~\cite{ninkovic_2025}:
\begin{equation}
\label{loss_AE}
\mathcal{L}_{\mathrm{UEP}} =
\lambda \sum_{k \in \mathcal{K}_{\mathrm{imp}}} \|\hat{z}_k - z_k\|^2
+ (1-\lambda) \sum_{k \notin \mathcal{K}_{\mathrm{imp}}} \|\hat{z}_k - z_k\|^2,
\end{equation}
where $\mathcal{K}_{\mathrm{imp}} \subset \{1,\dots,K\}$ denotes
task-critical latent components and $\lambda\in(0.5,1)$
controls the prioritization strength.
Prior AE-based UEP methods rely on predefined or reconstruction-driven
criteria to determine $\mathcal{K}_{\mathrm{imp}}$~\cite{ninkovic_2025,fayaz_2024},
without explicitly considering the downstream task~\cite{Liu_2026}.  In contrast, the
proposed framework determines $\mathcal{K}_{\mathrm{imp}}$ dynamically according
to task relevance. Thus, in this work, we preserve the weighted-
loss realization itself, but the contribution lies in the task-driven criterion for
selecting the protected latent components, which is equally
applicable to alternative UEP mechanisms.

\subsection{System Framework for Semantic Split Learning}
\label{sec:system_framework}

We consider a unified semantic split learning framework for edge--cloud
inference that integrates SL, semantic communication, and AE-based PHY design,
as illustrated in Fig.~\ref{fig:sys_mod}. At the edge device, input $X\in\mathbb{R}^N$ is processed by a semantic encoder
$f_{\mathrm{sem}}$, extracting  $Z=f_\mathrm{sem}(X)$.
A channel encoder $f_{\mathrm{ch}}$ then produces 
transmit signal $S=f_\mathrm{ch}(Z)$, so 
$f_{\mathrm{E}} = f_{\mathrm{ch}} \circ f_{\mathrm{sem}}$.

At the server, $\tilde{S}=\mathcal{C}(S)$, where $\mathcal{C}$  represents the physical communication channel, is
 processed by channel decoder $g_{\mathrm{ch}}$ to yield
$\hat{Z}=g_\mathrm{ch}(\tilde{S})$, and passed to semantic decoder $g_{\mathrm{sem}}$ for the
task estimate $\hat{T}=g_\mathrm{sem}(\hat{Z})$, i.e.,
$f_\mathrm{S} = g_{\mathrm{sem}} \circ g_{\mathrm{ch}}$. The overall
end-to-end system is:
\begin{equation}
\hat{T}
=
g_{\mathrm{sem}}
\!\left(
g_{\mathrm{ch}}
\!\left(
\mathcal{C}
\!\left(
f_{\mathrm{ch}}
\!\left(
f_{\mathrm{sem}}(X)
\right)
\right)
\right)
\right).
\end{equation}
This separation enables explicit allocation of communication resources to task-relevant components, forming the basis for the proposed task-aware UEP.

\section{Proposed Semantic Split Learning Framework with Learning-Driven UEP}

\subsection{Semantic Model Architecture and Split Design}
\label{sec:arch_split}

We consider a time-series IoT sensing task in which an edge device performs local feature extraction and collaborates with a server for inference. The semantic component is implemented as a two-layer 
long short-term memory (LSTM) network followed by a  fully connected (FC) layer. The first LSTM layer, deployed at the edge,  serves as
$f_{\mathrm{sem}}$, while the second LSTM and  FC
 serve as
$g_{\mathrm{sem}}$.
This split enables the edge  to extract task-relevant
 features locally. Each LSTM layer employs $K=10$ hidden units.
The FC layer maps the final hidden state to a
scalar output $\hat{T}$.
For benchmarking, a centralized baseline is considered,
in which all semantic components execute at the server without edge-side
preprocessing. To verify that the proposed approach is not tied to a specific latent dimension, we additionally evaluate $K\in\{20,40\}$; results are reported in
Section~\ref{sec:analysis} (Table~\ref{tab:combined}, top part). %and confirm consistent
%performance trends. %indicating that the observed gains are not an artifact of
%the low-dimensional bottleneck at $K=10$.}
%\Rtwo{To verify that the proposed approach is not tied to a specific latent
%dimension, we additionally evaluate $K\in\{20,40\}$ (Table~\ref{tab:K}).
%The results show consistent task performance trends, indicating that the
%observed gains are not an artifact of the low-dimensional bottleneck at $K=10$.}

\subsection{AE-Based PHY Layer and Communication Interface}
\label{sec:phy_interface}

The semantic encoder output $Z=f_{\mathrm{sem}}(X)$ is processed by channel
encoder $f_{\mathrm{ch}}$, yielding transmit signal $S = f_{\mathrm{ch}}(Z)$,
normalized to satisfy an average power constraint. The channel encoder is
 a FC network with a single hidden layer
of 10 neurons and batch normalization. The output dimension $n$
 controls communication bandwidth.
We use two transmit dimensions, $n\in\{5, 15\}$.

The channel is modeled as a real-valued Rayleigh block-fading channel:
$\tilde{S}=\mathcal{C}(S)=\mathbf{h}\odot S+\boldsymbol{\eta}$,
where $\mathbf{h}\in\mathbb{R}^n$ has entries
$h_\ell\sim\mathrm{Rayleigh}(1)$ and
$\boldsymbol{\eta}\in\mathbb{R}^n$ is AWGN with
$\eta_\ell\sim\mathcal{N}(0,\sigma^2)$.
Fading coefficients vary independently across blocks.
This standard abstraction~\cite{gunduz_2023} enables controlled evaluation of the proposed method.
Unlike conventional receivers, the proposed AE-based PHY layer does not require explicit CSI estimation; instead, $g_{\mathrm{ch}}$ learns an implicit inverse of the channel statistics through end-to-end training~\cite{OShea_2017}, with robustness determined by the match between training and deployment channel distributions.
Practical extensions to distributional channel mismatch, hardware impairments, and comparisons with standardized channel coding schemes under latency constraints are left for future work. Notably, the proposed MI-guided prioritization operates at the representation level and is agnostic to the specific transmission scheme.

At the server, $\tilde{S}$ is processed by channel decoder $g_{\mathrm{ch}}$,
producing $\hat{Z}=g_{\mathrm{ch}} (\tilde{S})$. 
Although there is no explicit power allocation layer, reliability
differentiation emerges through the \emph{learned representation geometry}: by
assigning larger reconstruction weights to selected latent components in the
loss, the optimizer maps those components to channel configurations with higher
effective signal energy~\cite{ninkovic_2025}. %This is empirically confirmed in
%Section~\ref{sec:analysis} (Table~\ref{tab:combined}, bottom part).
%\Rtwo{Although there is no explicit power allocation layer, reliability
%differentiation emerges through the \emph{learned representation geometry}: by
%assigning larger reconstruction weights to selected latent components in the
%loss, the optimizer maps those components to channel configurations with higher
%effective signal energy~\cite{Ninkovic_2021}. This is empirically confirmed in
%Table~\ref{tab:symbol_energy}, which shows that MI-guided UEP assigns $+21\%$
%more symbol energy to the top-ranked components compared to the EEP baseline.}

\subsection{Learning-Driven UEP and End-to-End Training}
\label{sec:training_sec}

UEP is realized through a learning-driven mechanism that 
prioritizes task-relevant latent components.
The importance of individual components is identified during
training based on their task contribution. The $L$ most task-relevant components of $Z$ ($L < K$, $L$ is a
hyperparameter) are identified using MI between $Z$ and target
$T$. A neural discriminator $f_\theta(Z,T)$ (two-layer FC  with
ReLU) approximates $I(Z;T)$ via the Donsker--Varadhan (DV) bound~\cite{dv_bound}:
\begin{equation}
\label{eq:mi_dv}
\hspace{-0.197cm}\hat{I}(Z;T) =
\mathbb{E}_{P(Z,T)}[f_\theta(Z,T)]
- \log \mathbb{E}_{P(Z)P(T)}\!\left[e^{f_\theta(Z,T')}\right].
\end{equation}
This criterion admits an information-theoretic interpretation. Since
$I(Z;T)=H(T)-H(T|Z)$ and $H(T)$ is independent of $Z$, we have
$\partial I(Z;T)/\partial z_k=-\partial H(T|Z)/\partial z_k$. Consequently,
$|\partial\hat I/\partial z_k|$ provides a local, differential surrogate for
the sufficient-statistics criterion underlying the IB principle. Under the
first-order approximation,
$\Delta\hat I(Z;T)\approx\sum_k(\partial\hat I/\partial z_k)\Delta z_k$,
components with larger gradient magnitudes correspond to perturbation
directions that most strongly affect task-relevant information. This
interpretation is most reliable when latent components are not highly
redundant,  and perturbations
remain within the local linear regime. Importantly, the DV estimate is used only for relative ranking rather than absolute MI
estimation, making the prioritization largely insensitive to estimator bias.
The stability of this mechanism is empirically validated in
Section~\ref{sec:analysis} (Fig.~\ref{fig:mi_conv}).

%\Rone{The stability of this mechanism was empirically validated across batch
%sizes $\{32,64,128,512\}$:
%Fig.~\ref{fig:mi_conv} shows that $\hat{I}(Z;T)$ converges stably for all
%batch sizes, with batch~64 closely following the smoother trajectories of
%larger batches. Also, the mean gradient standard deviation drops from
%$0.119$ at batch~$32$ to $0.051$ at batch~$64$ (a $>60\%$ reduction), and the
%Top-$5$ ranking consistency exceeds $88\%$ for all batch sizes, confirming
%that the induced ranking is not driven by stochastic fluctuations of the DV
%estimator.
%}

Per batch, the gradient of $\hat{I}(Z;T)$ w.r.t. $Z$ is
computed, and the $L$ components with the largest gradient magnitudes are
marked as task-critical.
The protection mechanism is fully learned and integrated: identification
of $\mathcal{K}_{\mathrm{imp}}$ and the update of the channel encoder/decoder
are performed in a single, unified backward pass, with no separate or manually
configured stage.

The AE reconstruction loss follows Eq.~\eqref{loss_AE}, with
$\mathcal{K}_{\mathrm{imp}}$ ($|\mathcal{K}_{\mathrm{imp}}|=L$) determined
 adaptively  using
MI-based ranking.

\textit{Training procedure:} All components $f_{\mathrm{sem}}$,
$f_{\mathrm{ch}}$, $g_{\mathrm{ch}}$, $g_{\mathrm{sem}}$, and $f_\theta$ are
trained jointly. The overall
loss is:
\begin{equation}
\label{eq:total_loss}
\mathcal{L}_{\text{total}} =
\mathcal{L}_{\text{task}}
+ \mathcal{L}_{\mathrm{UEP}}
+ \mathcal{L}_{\mathrm{MI}},
\end{equation}
where $\mathcal{L}_{\text{task}} = \mathbb{E}[(\hat{T} - T(X))^2]$
and $\mathcal{L}_{\mathrm{MI}}$ is the negative DV bound for the MI estimator.
Training uses Adam~\cite{adam} with $\alpha=9\times10^{-4}$, batch size 64,
and 100 epochs.
Per-batch SNR is sampled uniformly from $[5,10]$~dB.
The discriminator $f_\theta$ introduces only a modest training overhead:
the MI estimator requires $\approx7.1\times10^4$ floating point operations (FLOPs) per sample compared to
the base model complexity of $\approx8.5\times10^4$ FLOPs, corresponding to an
$11.7\%$ increase in processing time per batch (measured wall-clock training time). For on-device personalization scenarios, the
discriminator's $\approx3.6\times10^4$ parameters ($\approx$144~kB at FP32)
impose a modest memory footprint, with retraining overhead remaining in the
same $\sim$$12\%$ range; a detailed hardware-level characterization is left
for future work. This overhead is incurred only
during offline training; the deployed edge model retains the same runtime
footprint as the baseline SL architecture, imposing no additional computation
on IoT devices at inference time. 

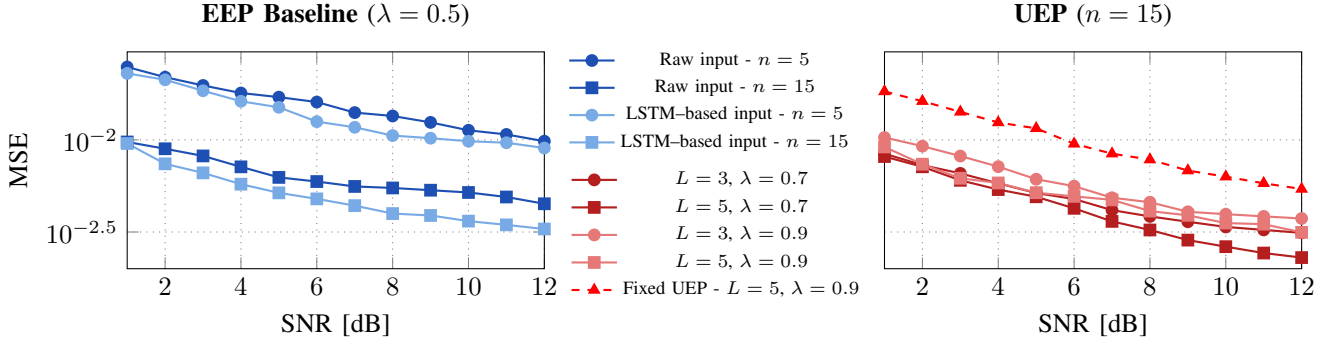
\begin{figure*}[t]
\centering
\begin{tikzpicture}
  \begin{groupplot}[
    group style={
      group size=2 by 1,
      horizontal sep=4.5cm,
      vertical sep=1.1cm,
    },
    width=7.1cm,
    height=4.45cm,
    ymode=log,
    xmin=1,
    xmax=12,
    grid=both,
    xlabel={SNR [dB]},
  ]

  \nextgroupplot[title={\textbf{EEP Baseline} ($\lambda=0.5$)},
    ylabel={MSE},
    legend style={at={(1.4,1.05)},anchor=north,font=\scriptsize,draw=none},
    ymin=0.002, ymax=0.03]
  \addplot[color=EEPdark,thick,mark=*,mark options={solid}]
    table[x={x},y={y}]{./fig_pred/5_raw.txt};
    \addlegendentry{Raw input - $n=5$}
  \addplot[color=EEPdark,thick,mark=square*,mark options={solid}]
    table[x={x},y={y}]{./fig_pred/15_raw.txt};
    \addlegendentry{Raw input - $n=15$}
  \addplot[color=EEPlight,thick,mark=*,mark options={solid}]
    table[x={x},y={y}]{./fig_pred/5_lstm.txt};
    \addlegendentry{LSTM--based input - $n=5$}
  \addplot[color=EEPlight,thick,mark=square*,mark options={solid}]
    table[x={x},y={y}]{./fig_pred/15_lstm.txt};
    \addlegendentry{LSTM--based input - $n=15$}

  \nextgroupplot[title={\textbf{UEP} ($n=15$)},
    legend style={at={(-0.4,0.5)},anchor=north,font=\scriptsize,draw=none},
    ymin=0.002, ymax=0.03, ymajorticks=false]
  \addplot[color=UEPdark,thick,mark=*,mark options={solid}]
    table[x={x},y={y}]{./UEP/n_15_MIB_3_MSE.txt};
    \addlegendentry{$L=3$, $\lambda=0.7$}
  \addplot[color=UEPdark,thick,mark=square*,mark options={solid}]
    table[x={x},y={y}]{./UEP/n_15_MIB_5_MSE.txt};
    \addlegendentry{$L=5$, $\lambda=0.7$}
  \addplot[color=UEPlight,thick,mark=*,mark options={solid}]
    table[x={x},y={y}]{./UEP/n_15_MIB_3_MSE_09.txt};
    \addlegendentry{$L=3$, $\lambda=0.9$}
  \addplot[color=UEPlight,thick,mark=square*,mark options={solid}]
    table[x={x},y={y}]{./UEP/n_15_MIB_5_MSE_09.txt};
    \addlegendentry{$L=5$, $\lambda=0.9$}
  \addplot[color=red,thick,dashed,mark=triangle*,mark options={solid}]
    table[x={x},y={acc}]{./UEP_no_MI/09_01_01.txt};
    \addlegendentry{Fixed UEP - $L=5$, $\lambda=0.9$}
  \end{groupplot}
\end{tikzpicture}
\caption{Prediction MSE, $\mathbb{E}\!\left[(\hat{T}-x_{t+1})^2\right]$,
vs.\ SNR under Rayleigh fading -- Left: EEP baseline; Right: UEP ($n=15$).}
\label{Fig_pred}
\end{figure*}

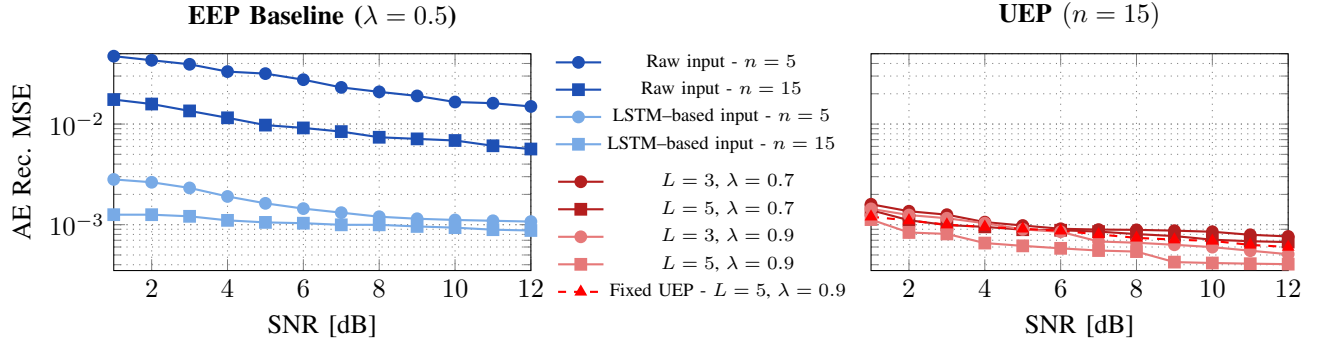
\begin{figure*}[t]
\centering
\begin{tikzpicture}
  \begin{groupplot}[
    group style={
      group size=2 by 1,
      horizontal sep=4.5cm,
      vertical sep=1.1cm,
    },
    width=7.1cm,
    height=4.45cm,
    ymode=log,
    xmin=1,
    xmax=12,
    grid=both,
    xlabel={SNR [dB]},
  ]

  \nextgroupplot[title={\textbf{EEP Baseline ($\lambda=0.5$)}},
    ylabel={AE Rec. MSE},
    legend style={at={(1.4,1.05)},anchor=north,font=\scriptsize,draw=none},
    ymin=0.00035, ymax=0.05]
  \addplot[color=EEPdark,thick,mark=*,mark options={solid}]
    table[x={x},y={y}]{./fig_AE/5_raw.txt};
    \addlegendentry{Raw input - $n=5$}
  \addplot[color=EEPdark,thick,mark=square*,mark options={solid}]
    table[x={x},y={y}]{./fig_AE/15_raw.txt};
    \addlegendentry{Raw input - $n=15$}
  \addplot[color=EEPlight,thick,mark=*,mark options={solid}]
    table[x={x},y={y}]{./fig_AE/5_lstm.txt};
    \addlegendentry{LSTM--based input - $n=5$}
  \addplot[color=EEPlight,thick,mark=square*,mark options={solid}]
    table[x={x},y={y}]{./fig_AE/15_lstm.txt};
    \addlegendentry{LSTM--based input - $n=15$}

  \nextgroupplot[title={\textbf{UEP $(n=15)$}},
    legend style={at={(-0.4,0.5)},anchor=north,font=\scriptsize,draw=none},
    ymin=0.00035, ymax=0.05, ymajorticks=false]
  \addplot[color=UEPdark,thick,mark=*,mark options={solid}]
    table[x={x},y={y}]{./UEP/n_15_MIB_3_AE.txt};
    \addlegendentry{$L=3$, $\lambda=0.7$}
  \addplot[color=UEPdark,thick,mark=square*,mark options={solid}]
    table[x={x},y={y}]{./UEP/n_15_MIB_5_AE.txt};
    \addlegendentry{$L=5$, $\lambda=0.7$}
  \addplot[color=UEPlight,thick,mark=*,mark options={solid}]
    table[x={x},y={y}]{./UEP/n_15_MIB_3_AE_09.txt};
    \addlegendentry{$L=3$, $\lambda=0.9$}
  \addplot[color=UEPlight,thick,mark=square*,mark options={solid}]
    table[x={x},y={y}]{./UEP/n_15_MIB_5_AE_09.txt};
    \addlegendentry{$L=5$, $\lambda=0.9$}
  \addplot[color=red,thick,dashed,mark=triangle*,mark options={solid}]
    table[x={x},y={rec}]{./UEP_no_MI/09_01_01.txt};
    \addlegendentry{Fixed UEP - $L=5$, $\lambda=0.9$}
  \end{groupplot}
\end{tikzpicture}
\caption{AE reconstruction MSE, $\|Z-\hat{Z}\|_2^2$, vs.\ SNR under Rayleigh
fading -- Left: EEP baseline; Right: UEP ($n=15$).}
\label{Fig_ae}
\end{figure*}

\section{Performance Evaluation}
\subsection{Experimental Setup}
\label{sec_evaluation}

The proposed framework is evaluated using data from a real-world IoT
environmental monitoring system on the Danube river near Novi Sad, Serbia.
The dataset consists of 3{,}264 daily measurements over nine years
(2013--2022), including dissolved oxygen, temperature, pH, and conductivity. The prediction task  forecasts daily dissolved oxygen levels.
Each input  consists of $N=20$ historical measurements, with the
target the next-day value.
Min--max scaling to $[-1,1]$ is applied.
The dataset is split chronologically into 70\% training/30\%
testing.

\subsection{Numerical Results}
\label{sec:performance_comparison}

We evaluate the impact of semantic preprocessing, latent dimension $n$, and
UEP on task-level prediction and representation robustness over a Rayleigh
block-fading channel. EEP corresponds to $\lambda=0.5$ in
Eq.~\eqref{loss_AE} and MI part is neglected.
As an additional UEP baseline, we consider a fixed prioritization scheme
in which the first five semantic components are predefined as
important for the case $n=15$.

\textbf{Task-level prediction performance:}
Fig.~\ref{Fig_pred} reports prediction MSE.
Under EEP (left panel), semantic preprocessing consistently improves prediction
accuracy across all SNRs.
For both $n\in\{5,15\}$, the LSTM encoder outperforms raw input
transmission, highlighting
the benefit of task-oriented feature extraction.
The trade-off between bandwidth and robustness to channel impairments is also
evident.
The LSTM-based SL approach achieves lower prediction MSE than the
centralized (raw input) baseline for both $n=5$ and $n=15$.

With UEP enabled (right panel), prediction performance improves further.
Protecting more components ($L=5$ vs. $L=3$) yields additional
gains.
The prioritization weight $\lambda$ plays a critical role: moderate
prioritization ($\lambda=0.7$) achieves the lowest prediction MSE, whereas
stronger prioritization ($\lambda=0.9$) degrades performance toward EEP, indicating that over-concentrating protection on a small subset is suboptimal.
MI-guided UEP significantly outperforms the fixed baseline  with consistently
lower prediction MSE across SNRs.

The task-relevance of the MI-gradient ranking was empirically confirmed
via a transmission ablation study: transmitting only the top-$5$ ranked
components yields MSE $= 0.00382$, vs.\ $0.00741$ for random selection and
$0.014$ for bottom-$5$ (at SNR=10 dB) -- a $>$$4\times$ gap. A perturbation analysis shows that injecting noise into the top-ranked components ($\sigma=0.1$) increases MSE to 0.00890, compared with 0.00627 for the bottom-ranked components, confirming that the MI-gradient criterion correctly identifies task-critical latent directions. Sensitivity analysis over $L\in\{1,\ldots,10\}$ at
$\mathrm{SNR}\in\{5,10,12\}$~dB shows a broad optimum around $L=5$, with
near-optimal performance for $L\in\{4,6\}$ across all tested SNRs, indicating
that per-SNR retuning is unnecessary.

\textbf{Representation reconstruction performance:}
Fig.~\ref{Fig_ae} shows AE reconstruction MSE.
Under EEP (left panel), semantic preprocessing improves reconstruction across all SNRs and maintains competitive performance even under reduced dimensionality.
With UEP (right panel), reconstruction accuracy improves across the full SNR range.
Unlike task-level performance, reconstruction error continues to benefit from
stronger prioritization ($\lambda=0.9$), revealing a divergence between
semantic fidelity and end-task accuracy.
That fixed UEP achieves similar reconstruction MSE to MI-guided UEP is
consistent with the proposed approach: both methods protect the same number $L$
of components with the same $\lambda$, so the aggregate
reconstruction loss is similar regardless of \emph{which} components are
protected. The reconstruction metric cannot
distinguish this; task MSE can. The superiority of MI-guided UEP is therefore
visible only in task-level performance (Fig.~\ref{Fig_pred}, right),
which is the metric that matters for semantic inference.

The divergence between reconstruction MSE and task MSE reflects a fundamental property of task-oriented systems: minimizing reconstruction error does not guarantee preservation of the task-relevant information required for predicting $T$~\cite{shao_2024}. Channel perturbations can disrupt task-critical latent components while leaving the average reconstruction error largely unchanged, resulting in disproportionate degradation of task accuracy. The proposed MI-based prioritization directly addresses this issue by protecting the components that contribute most to $I(Z;T)$. Together, the results show that semantic preprocessing improves robustness, while UEP provides targeted protection, highlighting their complementary roles and the importance of tuning UEP to the end task.

%The divergence between reconstruction MSE and task MSE reflects a
%fundamental property of task-oriented systems: minimizing MSE does not
%guarantee preservation of the sufficient statistics of $X$ for predicting
%$T$~\cite{shao_2024}. Channel perturbations may disrupt specific
%task-critical latent components while leaving the average reconstruction error
%largely unchanged, leading to disproportionate degradation in task accuracy.
%The proposed MI-based prioritization directly addresses this by protecting the
%components most relevant to $I(Z;T)$. These results demonstrate that semantic preprocessing (improves robustness) and
%UEP (enables targeted  protection) provide complementary gains.
%The divergence between reconstruction fidelity and prediction
%accuracy confirms that minimizing representation error alone is insufficient, and underscores the importance of 
%tuning UEP parameters regarding the end task.

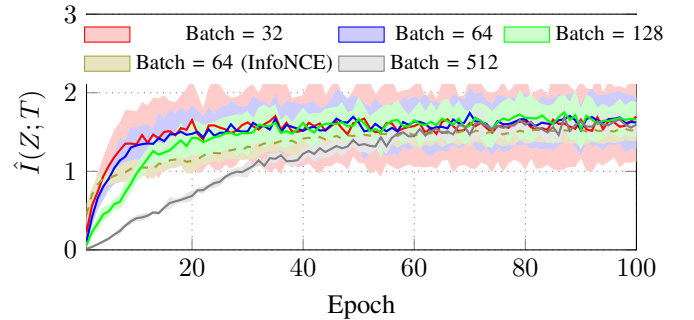
\begin{figure}[ht]
    \centering
\begin{tikzpicture}
\begin{axis}[
    width=1\linewidth,
    height=4.7cm,
    xlabel={Epoch},
    ylabel={$\hat{I}(Z;T)$},
    legend pos=north west,
    legend columns=3,
    grid=both,
    ymin=0, ymax=3,
    xmin=1, xmax=100,
    legend style={
    at={(-0.02,1)}, font=\footnotesize,
    anchor=north west,
    draw=none,
    forget plot
}
]

% ---- Batch size 64 example ----

% Upper bound (mean + std)
\addplot[
    name path=upper32,
    draw=none, 
    forget plot
]
table[x expr=\coordindex+1,
y expr=\thisrow{MI_mean} + \thisrow{MI_std}
] {new_results/Rev1/est_MI_32.txt};

% Lower bound (mean - std)
\addplot[
    name path=lower32,
    draw=none,
    forget plot
]
table[x expr=\coordindex+1,
y expr=\thisrow{MI_mean} - \thisrow{MI_std}
] {new_results/Rev1/est_MI_32.txt};

% Shaded area
\addplot[
    fill=red!20,
    draw=none, 
    forget plot
]
fill between[
    of=upper32 and lower32
];

% Mean curve
\addplot[
    red,
    thick,
    forget plot
]
table[x expr=\coordindex+1,
y=MI_mean
] {new_results/Rev1/est_MI_32.txt};

\addplot[
    name path=upper64,
    draw=none, 
    forget plot
]
table[x expr=\coordindex+1,
y expr=\thisrow{MI_mean} + \thisrow{MI_std}
] {new_results/Rev1/est_MI_64.txt};

% Lower bound (mean - std)
\addplot[
    name path=lower64,
    draw=none,
    forget plot
]
table[x expr=\coordindex+1,
y expr=\thisrow{MI_mean} - \thisrow{MI_std}
] {new_results/Rev1/est_MI_64.txt};
% Shaded area
\addplot[
    fill=blue!20,
    draw=none,
    forget plot
]
fill between[
    of=upper64 and lower64
];

% Mean curve
\addplot[
    blue,
    thick,
    forget plot
]
table[x expr=\coordindex+1,
y=MI_mean
] {new_results/Rev1/est_MI_64.txt};

\addplot[
    name path=upper64,
    draw=none, 
    forget plot
]
table[x expr=\coordindex+1,
y expr=\thisrow{InfoNCEmean} + \thisrow{InfoNCEstd}
] {new_results/Rev1/InfoNCE.txt};

% Lower bound (mean - std)
\addplot[
    name path=lower64,
    draw=none,
    forget plot
]
table[x expr=\coordindex+1,
y expr=\thisrow{InfoNCEmean} - \thisrow{InfoNCEstd}
] {new_results/Rev1/InfoNCE.txt};
% Shaded area
\addplot[
    fill=olive!20,
    draw=none,
    forget plot
]
fill between[
    of=upper64 and lower64
];

% Mean curve
\addplot[
    olive!80,
    dashed,
    thick,
    forget plot
]
table[x expr=\coordindex+1,
y=InfoNCEmean
] {new_results/Rev1/InfoNCE.txt};

\addplot[
    name path=upper128,
    draw=none,
    forget plot
]
table[x expr=\coordindex+1,
y expr=\thisrow{MI_mean} + \thisrow{MI_std}
] {new_results/Rev1/est_MI_128.txt};

% Lower bound (mean - std)
\addplot[
    name path=lower128,
    draw=none,
    forget plot
]
table[x expr=\coordindex+1,
y expr=\thisrow{MI_mean} - \thisrow{MI_std}
] {new_results/Rev1/est_MI_128.txt};

% Shaded area
\addplot[
    fill=green!20,
    draw=none,
    forget plot
]
fill between[
    of=upper128 and lower128
];

% Mean curve
\addplot[
    green,
    thick,
    forget plot
]
table[x expr=\coordindex+1,
y=MI_mean
] {new_results/Rev1/est_MI_128.txt};

\addplot[
    name path=upper512,
    draw=none,
    forget plot
]
table[x expr=\coordindex+1,
y expr=\thisrow{MI_mean} + \thisrow{MI_std}
] {new_results/Rev1/est_MI_512.txt};

% Lower bound (mean - std)
\addplot[
    name path=lower512,
    draw=none,
    forget plot
]
table[x expr=\coordindex+1,
y expr=\thisrow{MI_mean} - \thisrow{MI_std}
] {new_results/Rev1/est_MI_512.txt};

% Shaded area
\addplot[
    fill=gray!20,
    draw=none,
    forget plot
]
fill between[
    of=upper512 and lower512
];

% Mean curve
\addplot[
    gray,
    thick
]
table[x expr=\coordindex+1,
y=MI_mean
] {new_results/Rev1/est_MI_512.txt};
\addlegendimage{area legend, fill=red!20, draw=red}
\addlegendentry{Batch  = 32}

\addlegendimage{area legend, fill=blue!20, draw=blue}
\addlegendentry{Batch = 64}

\addlegendimage{area legend, fill=green!20, draw=green}
\addlegendentry{Batch = 128}

\addlegendimage{area legend, fill=olive!20, draw=olive!80}
\addlegendentry{Batch = 64 (InfoNCE)}
\addlegendimage{area legend, fill=gray!20, draw=gray}
\addlegendentry{Batch = 512}

% ---- Repeat for other batch sizes ----

\end{axis}
\end{tikzpicture}
\caption{Estimated MI $\hat{I}(Z;T)$ during training for different batch sizes and estimator types - Mean (line) $\pm$ std (shaded).}
\label{fig:mi_conv}
\end{figure}

\subsection{Analysis of the Proposed UEP Mechanism}
\label{sec:analysis}
 
\textbf{DV estimator stability:}
Fig.~\ref{fig:mi_conv} shows $\hat{I}(Z;T)$ during training for batch sizes
$\{32,64,128,512\}$ (mean $\pm$ std).
The estimate converges stably across all configurations, with batch~64 closely
tracking the smoother trajectories of larger batches.
The mean gradient standard deviation drops from $0.119$ at batch~32 to
$0.051$ at batch~64, and the Top-5 ranking consistency
exceeds $88\%$ for all batch sizes, confirming that the induced ranking is not
driven by stochastic fluctuations of the DV estimator. To verify robustness to estimator choice, we repeated training with an InfoNCE-based estimator~\cite{oord_2019} (Fig. \ref{fig:mi_conv}). Both estimators converge stably, produce the same dominant Top-5 ranking, and achieve nearly identical task MSE, confirming that the ranking is not estimator-specific.

\begin{figure}[t]
\centering
\begin{tikzpicture}
\begin{axis}[
    width=0.9\linewidth,
    height=4.7cm,
    ymode=log,
    xmin=1,
    xmax=12,
    ymin=0.0045,
    ymax=0.06,
    xlabel={SNR [dB]},
    ylabel={Prediction MSE},
    grid=both,
    legend style={
        font=\scriptsize,
        at={(0.98,1.03)},
        anchor=north east,
        draw=none
    }
]

% Raw input (n=15)
\addplot[
    color=EEPdark,
    thick,
    mark=x,
    mark options={solid},
]
table [x=x,y=y] {./new_results/Rev1/fig_pred/15_raw.txt};
\addlegendentry{Raw input}

% LSTM input (n=15)
\addplot[
    color=EEPlight,
    thick,
    mark=o,
    mark options={solid},
]
table [x=x,y=y] {./new_results/Rev1/fig_pred/15_lstm.txt};
\addlegendentry{LSTM-based input}

% UEP L=5 lambda=0.7
\addplot[
    color=UEPdark,
    thick,
    mark=square*,
    mark options={solid},
]
table [x=x,y=y] {./new_results/Rev1/UEP/n_15_MIB_5_MSE.txt};
\addlegendentry{UEP ($L=5,\lambda=0.7$)}

% UEP L=5 lambda=0.9
\addplot[
    color=UEPlight,
    thick,
    mark=*,
    mark options={solid},
]
table [x=x,y=y] {./new_results/Rev1/UEP/n_15_MIB_5_MSE_09.txt};
\addlegendentry{UEP ($L=5,\lambda=0.9$)}

\end{axis}
\end{tikzpicture}

\caption{Prediction MSE, $\mathbb{E}\!\left[(\hat{T}-x_{t+1})^2\right]$,
vs. SNR under Rayleigh fading - Amazon dataset ($n=15$).}
\label{Fig_pred_new}
\end{figure}
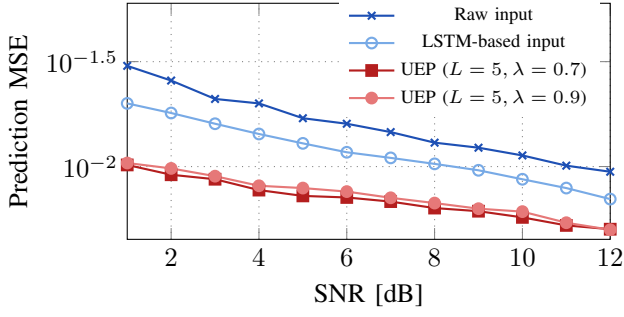

\textbf{Latent dimension ablation:}
Table~\ref{tab:combined} (top part) reports task MSE and FLOPs for $K\in\{10,20,40\}$.
Increasing $K$ yields only marginal accuracy improvements at significantly
higher complexity, and the relative gains of the proposed task-aware UEP
remain consistent across all values of $K$, confirming that the
observed gains originate from the task-aware UEP mechanism rather than
low-dimensional bottleneck effects.

%Table~\ref{tab:combined} (top part) reports task MSE and FLOPs for $K\in\{10,20,40\}$.
%Increasing $K$ yields only marginal accuracy improvements at significantly
%higher complexity, and the relative gains of the proposed task-aware UEP
%remain consistent across all values of $K$, confirming that the observed
%improvements are not an artifact of the low-dimensional bottleneck at $K=10$.

\textbf{PHY-layer differentiation:}
Table~\ref{tab:combined} (bottom part) reports the mean transmitted symbol energy
per component group after training with MI-guided UEP. The top-ranked components receive $+21\%$ more energy than the EEP
baseline, confirming that the weighted loss translates into genuine
PHY-layer reliability differentiation rather than merely a loss-level
effect. This learned redistribution approximates task-aware resource allocation but does not constitute a proof of globally optimal power allocation, which lies beyond the scope of this work.

\begin{table}[h]
\centering
\caption{(Top) Impact of $K$ on task MSE and FLOPs ($SNR=10$ dB). (Bottom) Symbol energy per component group under MI-guided UEP ($\lambda=0.7$, $n=15$).}
\label{tab:combined}
\begin{tabular}{lcc}
\toprule
 $K$ & \textbf{Task MSE} & \textbf{FLOPs}\\
\midrule
10  & $0.00382$ & $8.5\times10^4$ \\
20  & $0.00327$ & $14.2\times10^4$ \\
40  & $0.0029$  & $37.2\times10^4$ \\
\midrule
\textbf{Component group} & $\mathbb{E}[s_i^2]$ & \textbf{vs.\ EEP} \\
\midrule
Top-5 (protected) & $1.24$ & $+21\%$ \\
Bottom-5          & $0.86$ & $-16\%$ \\
EEP baseline      & $1.02$ & ---     \\
\bottomrule
\end{tabular}
\end{table}

%\begin{figure}[t]
%\centering
%\begin{tikzpicture}
%  	\begin{semilogyaxis}[width=0.9\columnwidth, height=4.7cm, 
%	legend style={at={(0.65,0.39)}, anchor= north,font=\footnotesize, 
%    draw=none, legend style={nodes={scale=1, transform shape}}},
%   	legend cell align={left},
%	legend columns=1,   	 
%   	x tick label style={/pgf/number format/.cd,
%   	set thousands separator={},fixed},
%   	y tick label style={/pgf/number format/.cd,fixed, precision=2, /tikz/.cd},
%   	xlabel={SNR [dB]},
%   	ylabel={Classification Accuracy},
%   	grid=major,   	
%   	xmin = 0, xmax = 10,
%   	ymin=0.78, ymax=0.98,
%   	line width=0.8pt,  	
%   	tick label style={font=\footnotesize},]
%        \addplot[blue, mark=o] 
%   	table [x={SNR}, y={MI_UEP}] {./new_results/Rev1/new_exp_class.txt};
%    \addlegendentry{MI-guided UEP - $L=5$, $\lambda=0.9$}
%    \addplot[red, mark=x] 
%   	table [x={SNR}, y={Fixed_UEP}] {./new_results/Rev1/new_exp_class.txt};
%   	\addlegendentry{Fixed UEP - $L=5$, $\lambda=0.9$}
%    \addplot[green!60!black, mark=*] 
%   	table [x={SNR}, y={EEP}] {./new_results/Rev1/new_exp_class.txt};
%   	\addlegendentry{EEP}
% 	\end{semilogyaxis}
%	\end{tikzpicture}
%	\vspace*{-1mm}
%\caption{Classification acc vs. SNR on the UCI (HAR) dataset.}
%\label{fig:HAR_placeholder}
%\end{figure}

\textbf{Generalization:}
The proposed framework was further evaluated on the Amazon stock dataset
(6,516 daily observations) for next-day \textit{Open} value prediction using
the previous $N=20$ observations (Fig.~\ref{Fig_pred_new}). MI-guided UEP consistently outperformed EEP
across all tested SNRs. We further evaluated the proposed framework on
the UCI HAR classification benchmark~\cite{HAR_dataset}. MI-guided UEP also consistently
outperformed both EEP and fixed UEP across all tested SNRs, demonstrating
robustness across both regression and classification tasks.

\section{Conclusion}
This letter proposes semantic SL for wireless edge--cloud
inference  with a learning-driven UEP mechanism
 at the split interface.
Task-aware prioritization of latent representations in an AE-based
PHY layer  enables adaptive reliability allocation
under fading channels.
Numerical results  show
consistent improvements over EEP and fixed-UEP
baselines across  SNRs.
Results on a second dataset and across latent dimensions confirm the
generality of the approach.
Future work will consider larger edge--cloud deployments, extension to multimodal collaborative settings, dynamic selection of the protection 
budget $L$
and integration with sparsification techniques.

\end{document}